\documentclass[
 reprint,
superscriptaddress,
 amsmath,amssymb,
 aps,%linenumbers,
superscriptaddress,
 amsmath,amssymb,
 aps,longbibliography
]{revtex4-1}

\usepackage{graphicx}% Include figure files
\usepackage{dcolumn}% Align table columns on decimal point
\usepackage{bm}% bold math
\usepackage{amsmath,amssymb,amstext,amsthm}
\usepackage{braket}
\usepackage{bbold}
\usepackage{bbm}
\usepackage{xcolor}
\usepackage{ gensymb }
\begin{document}

\preprint{APS/123-QED}
%\linenumbers
\setlength{\abovedisplayskip}{1pt}
%\title{Direct detection of optical quantum states by humans}
\title{Direct discrimination of structured light by humans}

\author{D. Sarenac}
\email{dsarenac@uwaterloo.ca}
%\affiliation{Transformative Quantum Technologies, University of Waterloo, Waterloo, ON, Canada, N2L3G1}
\affiliation{Institute for Quantum Computing, University of Waterloo,  Waterloo, ON, Canada, N2L3G1}
\author{C. Kapahi} 
\affiliation{Institute for Quantum Computing, University of Waterloo,  Waterloo, ON, Canada, N2L3G1}
\affiliation{Department of Physics, University of Waterloo, Waterloo, ON, Canada, N2L3G1}
\author{A. E. Silva} 
\email{a8silva@uwaterloo.ca}
\affiliation{School of Optometry and Vision, University of Waterloo, Waterloo, ON, Canada, N2L3G1}
\author{D. G. Cory}
\affiliation{Institute for Quantum Computing, University of Waterloo,  Waterloo, ON, Canada, N2L3G1}
\affiliation{Department of Chemistry, University of Waterloo, Waterloo, ON, Canada, N2L3G1}
\author{I. Taminiau}
\affiliation{Institute for Quantum Computing, University of Waterloo,  Waterloo, ON, Canada, N2L3G1}
\author{B. Thompson} 
\affiliation{School of Optometry and Vision, University of Waterloo, Waterloo, ON, Canada, N2L3G1}
\author{D. A. Pushin}
%\email{dmitry.pushin@uwaterloo.ca}
\affiliation{Institute for Quantum Computing, University of Waterloo,  Waterloo, ON, Canada, N2L3G1}
\affiliation{Department of Physics, University of Waterloo, Waterloo, ON, Canada, N2L3G1}

\date{\today}% It is always \today, today,

%\begin{abstract}
%\end{abstract}

\pacs{Valid PACS appear here}

%\section{\label{sec:level1}Introduction}

\begin{abstract}
%The human capability to directly perceive the manifestations of quantum phenomena remains a distinct curiosity. The ordered structure in nature that is associated with well defined quantum states is typically outside the domain of human interaction.In this work we 
We predict and experimentally verify an entoptic phenomenon through which humans are able to perceive and discriminate structured light with space-varying polarization. Direct perception and discrimination is possible through the observation of distinct profiles induced by the interaction between the polarization gradients in a uniform-intensity beam and the radially symmetric dichroic elements that are centered on the foveola in the macula of the human eye. A psychophysical study was conducted where optical states with coupled polarization and orbital angular momentum (OAM) were directed onto the retina of participants. The participants were able to correctly discriminate between two states, differentiated by OAM $=\pm7$, with an average success probability of $77.6\%$ (average sensitivity $d^\prime=1.7$, $t(9) = 5.9$, $p = 2\times 10^{-4}$). These results enable new methods of robustly characterizing the structure of the macula, probing retina signalling pathways, and conducting experiments with non-separable optical states and human detectors.
\end{abstract}
\maketitle

%\section{Introduction}

\begin{figure*}
\centering\includegraphics[width=\linewidth]{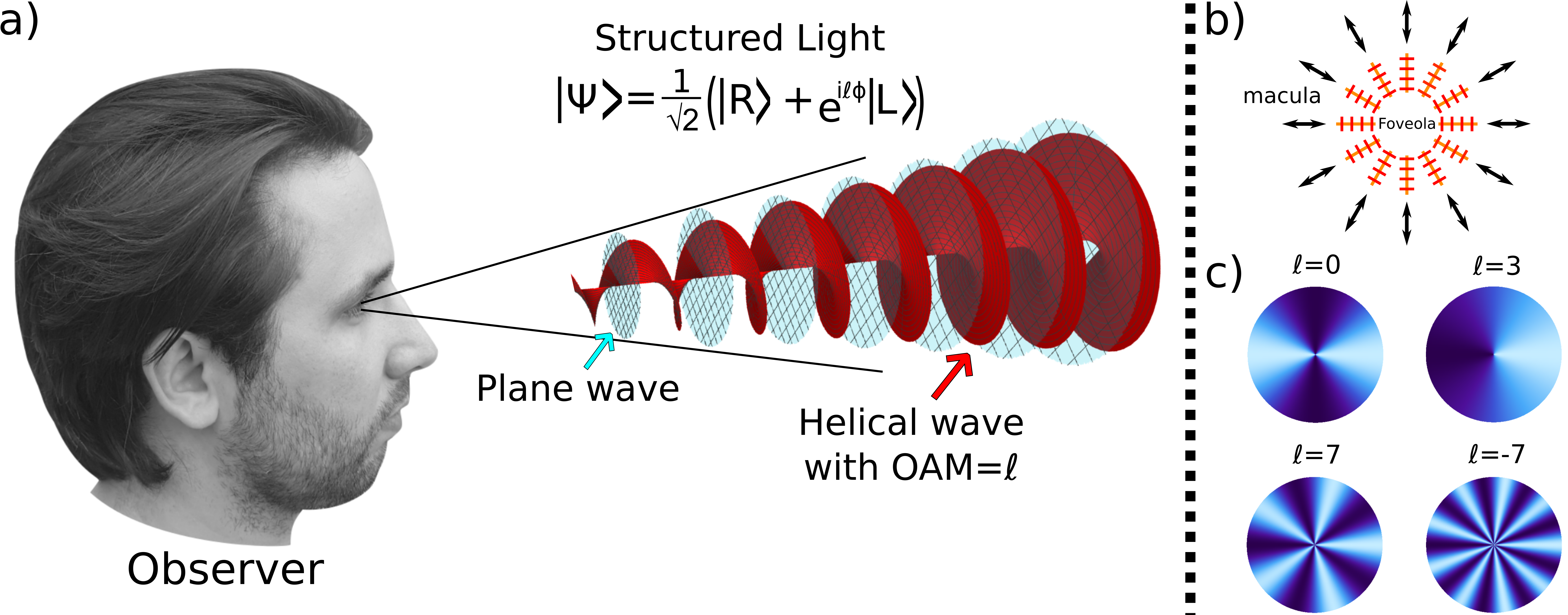}
\caption{a) Pictorial representation of structured light, composed of a coherent superposition of a planar right-circularly polarized state and a helical left-circularly polarized state, being directed onto the retina of an observer. The helical state carries orbital angular momentum (OAM) and its phase varies along the azimuthal coordinate $\phi$. b) In the macula of the human eye the macular pigment molecules (red) are bound to the radially oriented Henle fibers (orange) that surround the foveola. The radial symmetry of these dichroic elements (polarization filter direction shown by black arrows)  coincides with the symmetry of the polarization coupled OAM beams shown in a). c) Depending on the OAM of the helical beam the participant observes a different unique signature when looking in the vicinity of the beam's origin. Shown are the examples for $\ell=0,3,7,-7$. The number of azimuthal fringes that a human sees is equal to $\vert\ell-2\vert$. The $\ell=0$ case depicts the Haidinger's brush profile when horizontally polarized light is observed.}
 \label{fig:fig1}
\end{figure*}

The complexity and richness of the human visual system makes it a focus point of incredibly diverse research. Recent technological advances in optics have enabled accurate probing of human visual perception capabilities and limits. For example, it was shown that humans are able to detect single quanta of light, or photons, with a probability greater than chance~\cite{tinsley2016direct}. The long standing question being whether a single photon incident on a photoreceptor can be perceived~\cite{hecht1942energy,field2005retinal,phan2014interaction,holmes2015testing}. %Such studies pave the way for future experiments with humans as detectors in quantum related experiments [cite]. %Such studies provide intriguing answers on whether humans are inherently capable of detecting quantum aspects of nature and whether there are quantum functions in biology~\cite{lambert2013quantum,ball2011physics}. 
%In this letter we introduce and test completely new methods through which a person can observe and quantify optical quantum phenomena. The considered optical beams are in well defined quantum states.  with Pancharatnam-Berry geometrical phases which in the considered cases manifest themselves as spatially dependant polarization~\cite{pancharatnam1956generalized,berry1987adiabatic}. The introduced techniques then rely on the human ability to perceive and analyze the spatial polarization gradients of a uniform intensity monochromatic light beam. We conducted a study on whether people can discriminate between two optical quantum beams.
In this letter we explore a novel domain of human vision by experimentally verifying an entoptic phenomenon through which humans perceive and discriminate different forms of structured light with space-varying polarization. A pictorial representation is shown in Fig.~\ref{fig:fig1}a, where space-varying polarization profiles are realized via optical states with coupled polarization and orbital angular momentum (OAM). We demonstrate that the OAM modes of this particular form of structured light induce distinct entoptic images in humans. % The perceptual discrimination is based on the human ability to perceive polarization gradients of light, which may be derived from the well-known human ability to perceive the polarization of light. In this study structured light is confirmed to induce a new category of entoptic images.  

%Humans are generally oblivious of their ability to perceive the polarization of light. The entoptic phenomenon was first described by Austrian physicist Wilhelm Karl von Haidinger in 1844~\cite{haidinger1844ueber}.
The ability to perceive space varying polarization in structured light may be derived from an entoptic phenomenon through which humans can perceive the polarization state of light.% was first described by Austrian physicist Wilhelm Karl von Haidinger in 1844
~\cite{haidinger1844ueber}. %It is however speculated that the Vikings made use of this phenomenon for navigation~\cite{ropars2011depolarizer}. 
When viewing polarized light, a bowtie-like shape (known as ``Haidinger's brush'') appears in the central point of the visual field. Although the exact physiological origin of the Haidinger's brushes is not fully understood, the prominent theory suggests that the perception of Haidinger's brushes depends on the presence of radially symmetric dichroic elements that are centered on the foveola~\cite{horvath2004polarized}. This has led to studies on the use of Haidinger's brushes to assess central visual field dysfunction and age-related macular degeneration~\cite{forster1954clinical,naylor1955measurement}, and macular pigment density~\cite{muller2016perception}. 

%The ability to perceive polarization of light varies significantly among people, and for the same person the ability even varies for the left and the right eye~\cite{le2010polarization}. The clarity of the Haidinger's brushes is found to peak for blue light of $\approx460$ nm wavelength~\cite{bone1980role}, and the average polarization threshold for the detection of Haidinger's brushes is $\approx56\%$~\cite{temple2015perceiving}. 

The orientation of the Haidinger's brush depends on the polarization state of light. Linearly polarized light induces a brush oriented perpendicular to the polarization direction~\cite{haidinger1844ueber}, while the brush appears rotated $\approx45\degree$ clockwise (counter-clockwise) when viewing left (right) circularly polarized light~\cite{shurcliff1955haidinger}. %As circularly polarized light is not axis dependant, the orientation of the brush is independent of the rotation of the head. Hence looking at a point and rotating one’s head allows a person to determine the polarization state of light.
However, retinal adaptation causes Haidinger's brushes to disappear after a few seconds if the polarization direction relative to the eye does not change. It has been found that stable perception of the brushes is achieved when the linear-polarization source is rotated at $\approx1$ Hz~\cite{coren1971use}. One may observe the behaviour of Haidinger's brushes by looking at the light scattered in the clear sky $\approx90\degree$ from the sun~\cite{horvath2017celestial}. With some practice, a brush may be observed that points towards the sun. %One may verify this with a common smartphone which typically possesses a polarized screen; with some practice, Haidinger's brush may be observed when rotating the phone while viewing a blue picture on the screen.

In this study we consider perception of structured light with a polarization direction that varies across the beam. The general form of the transverse wavefunction of a spatially dependant optical state travelling along the z-direction is given by:

\begin{align}
	\ket{\Psi}=A_1(r,\phi)\ket{R}+e^{if(r,\phi)}A_2(r,\phi)\ket{L},
	\label{Eqn:PsiInGeneral}
\end{align}

\noindent where we have used the bra-ket notation for convenience, $(r,\phi)$ are the cylindrical coordinates, and $\ket{L}$ and $\ket{R}$ denote the left and right circular polarization. 

As shown in Fig.~\ref{fig:fig1}b, the macular pigment molecules (red) in the human macula are bound to the radially oriented Henle fibers (orange) that surround the foveola~\cite{horvath2004polarized}. The accepted model for the action of the macula on the incoming light is to treat it as a radial polarization filter~\cite{misson2003mueller,rothmayer2007nonlinearity,misson2018computational}, a concept dating back to Maxwell and Helmholtz~\cite{maxwell1850manuscript,von1925treatise}. The operator of the macula can therefore be expressed as: 

\begin{align}
	\ket{M}\bra{M}=\frac{1}{2}
	\begin{pmatrix}
1 & e^{-i2\phi} \\
e^{i2\phi} & 1
\end{pmatrix}
	\label{Eqn:eyeoperator}
\end{align}

Several theories have been put forward in order to account for the human perception of circularly polarized light. Good agreement is found when accounting for corneal birefringence that is uniformly along the visual axis~\cite{bour1991polarized,knighton2002linear,misson2003mueller}. The corresponding operator is given by:

\begin{align}
U_m=e^{i\alpha\hat{\sigma}_x}
\label{Eqn:CircularPhaseShift}
\end{align}

\noindent where $\hat{\sigma}_x$ is the Pauli operator. The clarity of the brush when viewing circularly polarized light is determined by the total amount of phase ($\alpha$) that the ocular birefringence induces, which is subject to individual variation~\cite{bour1991polarized,knighton2002linear,misson2003mueller}. The two operators of Eq.~\ref{Eqn:eyeoperator}~$\&$~\ref{Eqn:CircularPhaseShift} acting on a polarized light beam reproduce with good agreement the reported descriptions of the Haidinger's brushes.

The same equations can be used to predict how humans would perceive structured light and polarization gradients. It follows that the profile that a person would perceive when viewing an arbitrary structured light beam is given by:

\begin{align}
I=|\bra{M}U_m\ket{\Psi}|^2
\label{Eqn:observedIntensity}
\end{align}

\noindent where $\ket{\Psi}$ is given by Eq.~\ref{Eqn:PsiInGeneral}. The radial symmetry of the macula in the human eye coincides with the symmetry of polarization coupled OAM states. The eye operator in Eq.~\ref{Eqn:eyeoperator} possesses an $e^{i2 \phi}$ term, whereas OAM states are associated with a helical phase front which is described by the factor $e^{i\ell \phi}$ in the wave function, where $\phi$ is the azimuthal coordinate and $\ell$ is the OAM number.  The emergence of structured beams and OAM states in light~\cite{LesAllen1992}, electrons~\cite{bliokh2007semiclassical,uchida2010generation,mcmorran2011electron}, and neutrons~\cite{Dima2015,sarenac2016holography,Sarenac201906861} has revolutionized quantum technologies and enabled numerous applications in microscopy, quantum information processing protocols, material characterization, and manipulation of matter~\cite{rubinsztein2016roadmap,BarnettBabikerPadgett,mair2001entanglement,wang2012terabit,Andersen2006,he1995direct,friese1996optical,brullot2016resolving,Simpson:97,sarenac2018generation,schwarz2019talbot}. Here we extend the control of structured light to visual science applications.

\begin{figure*}
\centering\includegraphics[width=\linewidth]{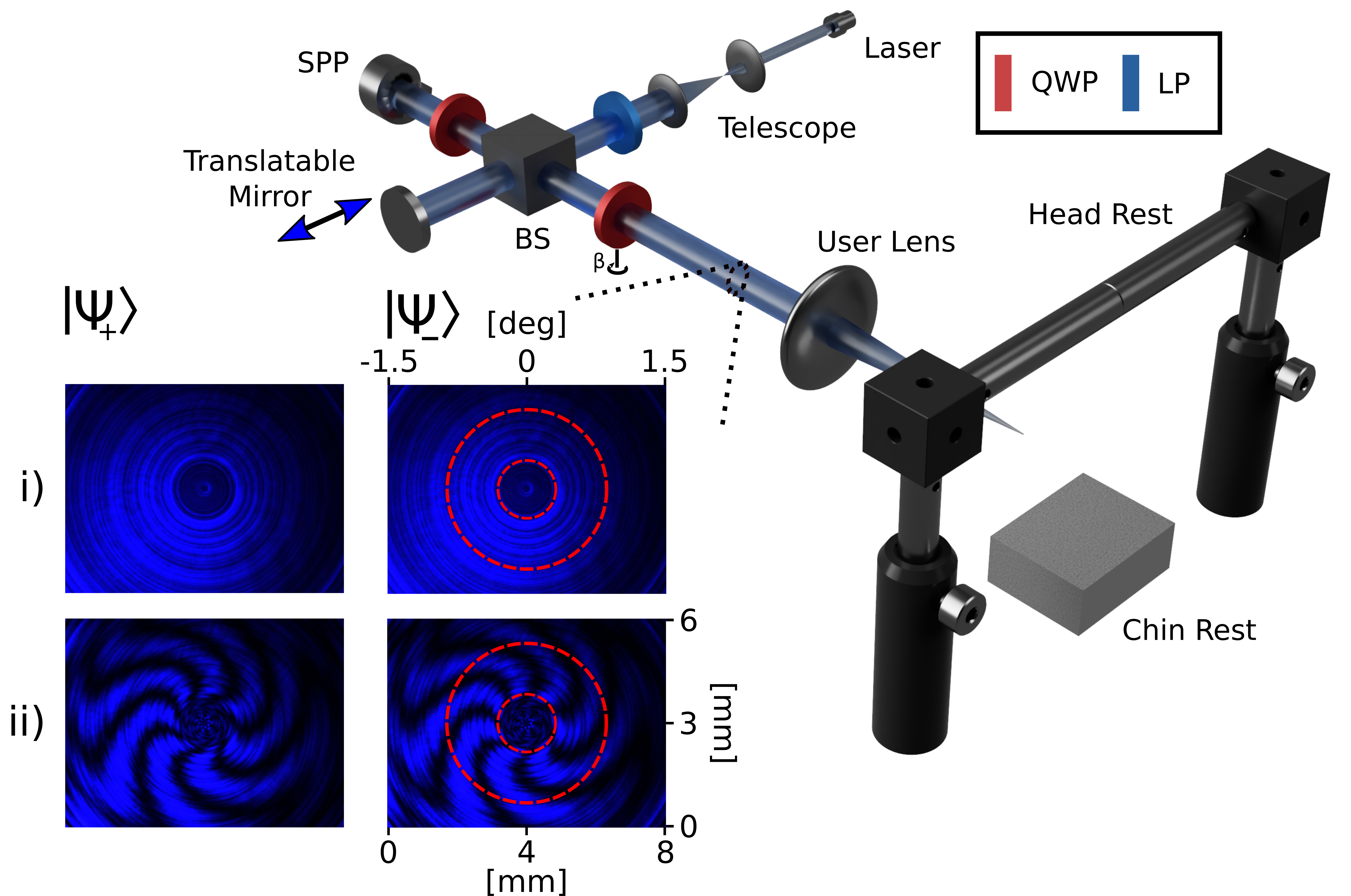}
\caption{Schematic of the experimental setup where a Michelson interferometer along with a spiral phase plate (SPP) and standard polarization optics components are used to prepare the structured light that is focused onto the retina of the participants in the study. For a complete description of the setup see Appendix A. Translating the mirror along the beam path direction varies $\theta(t)$ in Eq.~\ref{Eqn:psiSO}, while the two orientations of the outer quarter wave plate (QWP), $\beta\in[0,180\degree]$, correspond to the two states ($\ket{\Psi_+}$ and $\ket{\Psi_-}$) of Eq.~\ref{Eqn:psiSO}. i) The images observed by a CMOS camera placed before the user lens, for both $\ket{\Psi_+}$ and $\ket{\Psi_-}$.  It can be noted that azimuthal fringes are not present. The ring features are artifacts from SPP machining and they are equally present in both images. ii) The images observed by a CMOS camera placed before the user lens when a linear polarizer (LP) is placed in front of the camera. The seven azimuthal fringes correspond to the phase structure of $\ket{\Psi_+}$ and $\ket{\Psi_-}$, the only notable difference being the $180\degree$ azimuthal phase shift. The attenuators were removed to obtain the images shown in i) and ii) and the camera gain was correspondingly optimized. In the study the participants only observed beams shown in i), and the red circles bound the area ($\approx2\degree$ of field of vision) with good intensity and high quality phase structure that the participants were instructed to observe. The two simulated profiles of what the participants were expected to observe are shown in Fig.~\ref{fig:fig1}c under the labels ``$\ell=7$'' for $\ket{\Psi_+}$ and ``$\ell=-7$'' for $\ket{\Psi_-}$.}
 \label{fig:setup}
\end{figure*}

%We note that in our initial tests we did not observe a unique signature when looking at a unpolarized beam carrying OAM, or a difference in the Haidinger brush structure due to OAM being induced in a polarized beam. 
The schematic of the experimental setup is shown in Fig.~\ref{fig:setup}. The laser light was attenuated to $\textless1$ $\mu$W/mm$^2$ at the location of the observer in order to conform to the guidelines for laser exposure time outlined by the International Commission on Non-Ionizing Radiation ~\cite{international2000revision}. A spiral phase plate (SPP)~\cite{beijersbergen1994helical} was placed in one arm of a Michelson interferometer along with standard polarization components. The SPP in reflection mode induced OAM of $\ell=7$ for $\lambda=450$ nm light. The setup thus allowed us to prepare and switch between the following two states:

\begin{align}
	\ket{\Psi_\pm}=\frac{1}{\sqrt{2}}\left(\ket{R}
                +e^{  i\theta(t)}e^{ \pm i7\phi}\ket{L}\right),
	\label{Eqn:psiSO}
\end{align}
\noindent where $\theta(t)$ is a linear phase in time which acts to rotate the polarization profile of the beam, analogous to rotating the polarization direction of a beam to induce high clarity Haidinger's brush~\cite{coren1971use}. By translating the mirror along the beam propagation direction we varied $\theta(t)$ by $\approx2\pi/7$ rad$/$s. For a complete description of the parameters in the setup see Appendix A.  It follows from Eq.~\ref{Eqn:psiSO} that the left circular polarization state of $\ket{\Psi_\pm}$ carries an OAM of $\pm\ell$, and that the spatially dependant phase shift $e^{ i\ell\phi}$ manifests into a space-varying polarization profile. 

For the two states of Eq.~\ref{Eqn:psiSO} we can determine the profiles that a person would observe using Eq.~\ref{Eqn:observedIntensity}. The two simulated profiles are shown in Fig.~\ref{fig:fig1}c under the labels ``$\ell=7$'' for $\ket{\Psi_+}$ and ``$\ell=-7$'' for $\ket{\Psi_-}$. Furthermore, using Eq.~\ref{Eqn:observedIntensity} we can determine that a person would observe $\vert\ell-2\vert$ number of azimuthal fringes when viewing a beam described by Eq.~\ref{Eqn:psiSO}. This can also be deduced by noting that the eye operator in Eq.~\ref{Eqn:eyeoperator} possesses a $e^{ i2\phi}$ term. The azimuthal fringes arise from the interference of the two polarization states that are carrying different OAM. Therefore a person may discriminate between the two states of Eq.~\ref{Eqn:psiSO} by observing the number of azimuthal fringes: $\ket{\Psi_+}$ manifests 5 azimuthal fringes and $\ket{\Psi_-}$ manifests 9 azimuthal fringes.

%The parameters in the setup were optimized and tested until the authors D. S. and C. K. could perform the discrimination task with $>80\%$ success rate. 
To test the hypothesis that human observers can discriminate between the two states of Eq.~\ref{Eqn:psiSO}, a psychophysical study was conducted where randomly selected states (either $\ket{\Psi_+}$ or $\ket{\Psi_-}$) were presented and participants discriminated between the two states based on the number of azimuthal fringes that they observed. Several factors helped ensure that the number of azimuthal fringes was the only cue for discriminating the beams. The setup used the orientation of the outer QWP to change between $\ket{\Psi_+}$ and $\ket{\Psi_-}$ while keeping the same SPP configuration. This ensured that the circular machining features noticeable in Fig.~\ref{fig:setup}i were equally present in both cases. The studies were done without any ambient light and there was a screen (with a $1$ inch diameter hole for the laser light to travel through) before the user lens which blocked the view of the setup by the participant. Furthermore, the QWP whose orientation determined which state was being observed was motorized to make an equal amount of motion between each trial. For a complete description of the psychophysical procedure see Appendix C.

\begin{figure}
\centering\includegraphics[width=\linewidth]{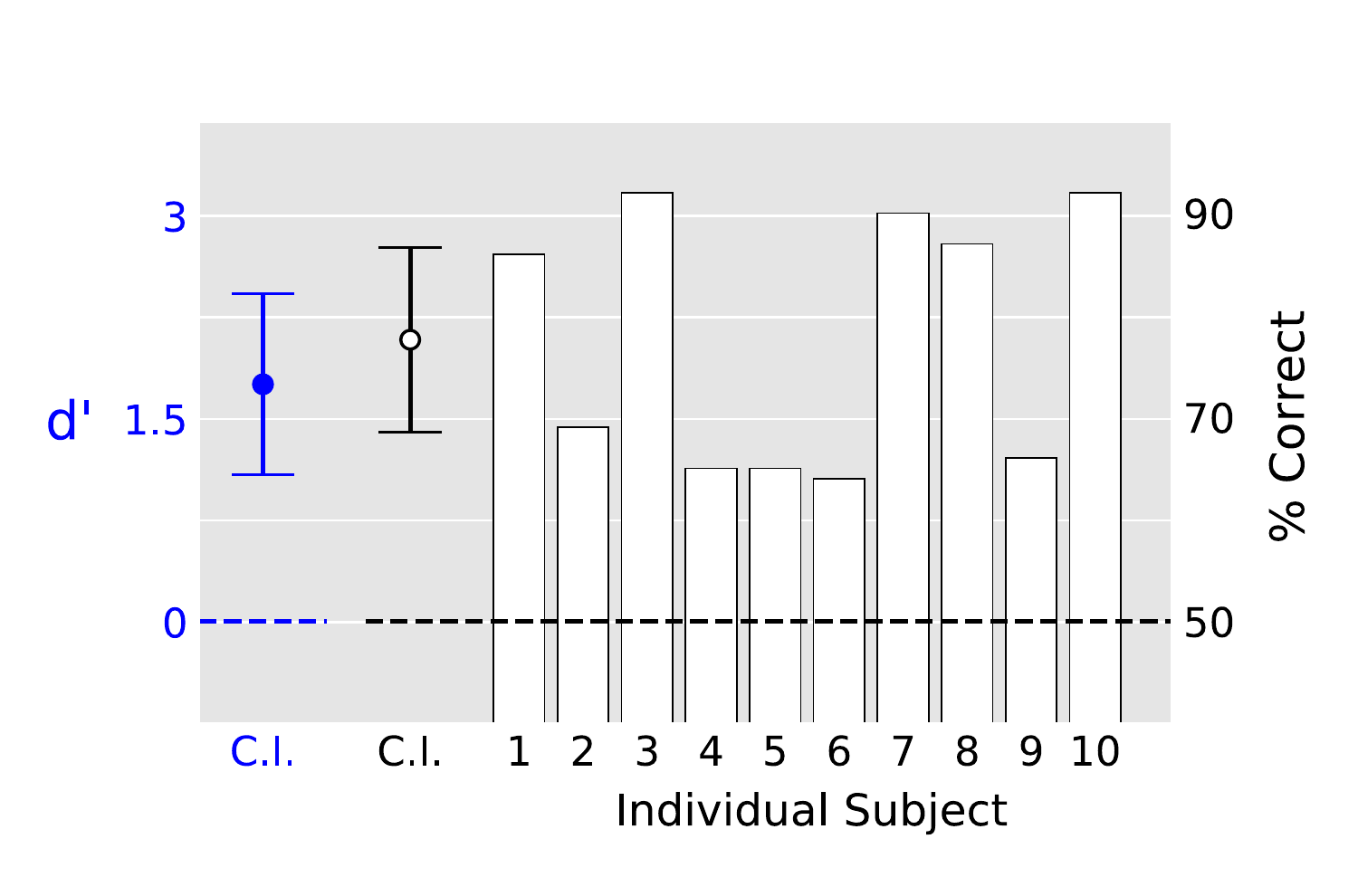}
\caption{Sensitivity and accuracy for the discrimination task. Participants each performed 100 trials over two sessions. The dashed line indicates chance performance. Open bars show individual participant performance. Circular symbols show group mean sensitivity (blue: left ordinate) and accuracy (black: right ordinate). Error bars show 95$\%$ confidence intervals. %Blue (left) axis: The 95$\%$ confidence interval for the average sensitivity scores. %This confidence interval is the estimated range exhibiting a 95 percent chance to encompass the true average sensitivity of the sampled population of participants. 
Participants were highly sensitive to the difference between both trial types, performing significantly better than chance.% (average $d^\prime=1.76$, $t(9) = 5.9$, $p = 2\times 10^{-4}$). Black axis: For better visualization, a corresponding 95 percent confidence interval for the average percent correct achieved is displayed alongside bars depicting individual-subject performance.
}
 \label{fig:results}
\end{figure}

After a brief familiarization period, the participants performed 100 random trials with structured light over two sessions on separate days. After viewing the stimulus, participants responded in one of two ways, responding ``many'' if they observed 9 rotating azimuthal fringes or responding ``fewer'' if they observed 5 rotating azimuthal fringes. Fig.~\ref{fig:results} shows the results for the ten participants who completed the study. There was no statistical difference between the results of session 1 and session 2, and therefore the data from both sessions were collapsed for the main analysis. 

Sensitivity $d^\prime$ and response bias $c$ were calculated for each participant. Percent correct is influenced by both a participant's ability to perform the task and the participant's response bias. However, $d^\prime$ is independent of response bias and is therefore a more accurate measure of performance when response bias is present~\cite{wickens2002elementary}. Data was analyzed using two-tailed, one-sample t-tests with 9 degrees of freedom against the null value of 0.

All participants achieved performance that is numerically above chance, and collectively they achieved good discrimination sensitivity, $d^\prime=1.7$, $t(9)=5.9$, \textit{p}-value: $p=2\times 10^{-4}$, corresponding to a mean accuracy of $77.6~\%$ correct. A significant response bias was also observed, $c=-0.2$, $t(9)=3.0$, $p=.02$, as participants responded ``many'' more often than ``fewer''. Fig.~\ref{fig:results} also suggests a bimodal distribution, where half of the participants achieved near-ceiling performance and the other half exhibited lower scores but remained above chance. There are no apparent explanations for this subdivision in terms of gender, age, or vision. We speculate that task performance is related to the various degrees of sensitivity that results from individual differences in the amount of ocular birefringence and the the organizational structure of the macula.

%There has been no biological function associated with the human perception of polarization (i.e. Haidinger's brushes). It is even more curious that this ability is suited towards perceiving structured light which in the case of interest does not occur in nature. Several examples can be found where the polarization varies in space, for example the light scattered in the sky $\approx90\degree$ from the sun~\cite{horvath2004polarized}. However, the authors could not find an example of structured light arising in nature that would enable the perception of profiles depicted in Fig.~\ref{fig:fig1}c as the primary prerequisite is coherent light. 

To the best of our knowledge these experiments provide the first exploration and confirmation of humans directly perceiving and discriminating structured light. Many follow up experiments are enabled given the recent advances in the control and manipulation of structured light. The setup in Fig.~\ref{fig:setup} can be improved by incorporating a spatial light modulator in place of the SPP. This would allow us to prepare arbitrary polarization gradients and test the psychophysical thresholds of human perception of polarization: the sensitivity distribution to a range of OAM numbers, individual differences in discrimination ability, and human sensitivity to other forms of structured light and polarization gradient patterns. Furthermore, optimizing the subjective clarity of the observed image allows us to determine the exact forms of  Eq.~\ref{Eqn:eyeoperator} $\&$ \ref{Eqn:CircularPhaseShift} for a particular person. The exact form of the operators is currently subject to debate~\cite{horvath2004polarized,misson2003mueller,rothmayer2007nonlinearity,misson2018computational}. 

Our follow up studies will also explore clinical applications of structured light perception. We speculate that structured light can be a highly sensitive probe of central visual field dysfunctions and age-related macular degeneration. Similar to fundus imaging with polarized light~\cite{hochheimer1982retinal,vannasdale2009determination} we can devise objective photographic tests with structured light. 

Given the non-separability of Eq.~\ref{Eqn:psiSO} an experiment can be conducted where the correlations between the two degrees of freedom (DOF), polarization and OAM, are confirmed with humans as detectors. The rotation of the profile that would be observed in the following two cases should be identical: phase shift on the OAM DOF (induced by rotating the SPP) and the phase shift on the polarization DOF (induced by a properly aligned birefringent material). \footnote{Note that the use of single photons instead of laser light would require an extremely bright single photon source as the intensity of the light at the location of the user lens was $\approx2$ nW.} 

\section{Acknowledgements}

This work was supported by the Canadian Excellence Research Chairs (CERC) program, the Natural Sciences and Engineering Research Council of Canada (NSERC) grants RGPIN$-2018-04989$, RPIN$-05394$, RGPAS$-477166$, the Collaborative Research and Training Experience (CREATE) program, and the Canada  First  Research  Excellence  Fund  (CFREF). The authors are thankful to Alex Mitrovic for his help with machining. D. A. P is thankful to Dusan Mirkovic for useful discussions.

\section{Contributions}

D. Sarenac postulated the idea. D. Sarenac, C. Kapahi, D. G. Cory, and D. A. Pushin designed and constructed the optical setup. I. Taminiau developed the spiral phase plate. D. Sarenac, C. Kapahi, A. E. Silva and D. A. Pushin optimized the setup according to the subjective perceptions of D. Sarenac and C. Kapahi. Then A. E. Silva and B. Thompson designed the psychophysical study. D. Sarenac, C. Kapahi, and A. E. Silva carried out the study. A. E. Silva and B. Thompson analyzed the results. All authors contributed to writing the paper.

\bibliography{OAM}

\clearpage

\section{APPENDIX}
%\section{Methods}
\subsection{Setup and Stimuli}

\begin{figure*}
\centering\includegraphics[width=.75\linewidth]{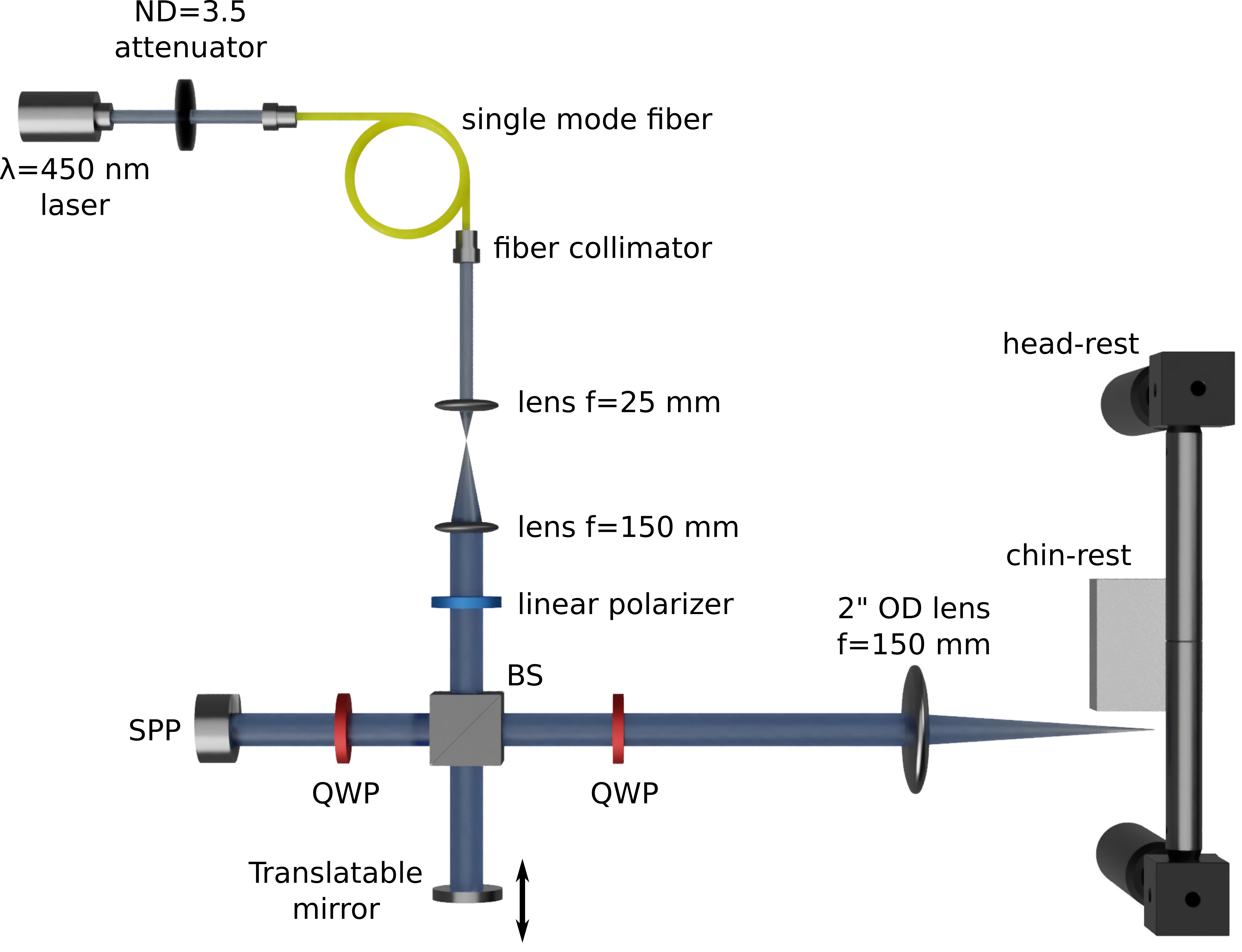}
\caption{Detailed schematic of the setup.}
 \label{fig:SuppFigSetup}
\end{figure*}

The detailed schematic of the setup is shown in Fig.~\ref{fig:SuppFigSetup}. For this experiment we adhere to the guidelines for laser exposure time outlined by the International Commission on Non-Ionizing Radiation Protection which state that the maximum permissible exposure for a human eye for blue wavelengths is 1 $\mu$W/mm$^2$~\cite{international2000revision}. Attenuators of ND=3.5 were placed after the laser in the setup and the intensity of light was $\approx 2.9$ nW before the user lens. It was confirmed that the power density of light near the focal spot (near the eye location) was well below the stated limit. The apparatus was approved for use with human participants by the University of Waterloo Ethical Review Board following an assessment by the University of Waterloo Safety Office. 

As the clarity of the Haidinger's brushes peaks for blue light of $\approx460$ nm wavelength, we used a diode laser with a central wavelength of 450 nm. A single mode optical fiber was used to clean the beam which was then expanded to 1.25 cm diameter via a 2 lens telescope system ($f_1= 25$ mm and $f_2= 150$ mm). The beam was then passed through a vertical polarizer. A Michelson interferometer was used to prepare the states of Eq.~\ref{Eqn:psiSO}. A beamsplitter first creates a coherent superposition of two paths. One of the paths is reflected by a mirror and the other path is reflected by a spiral phase plate (SPP). The SPP was generated out of 4N purity aluminum in an ultra precision machining center using custom diamond tooling. Temperature control was kept within $1\degree$C and form accuracy was limited by the thermal expansion of the aluminum due to any thermal drift. The SPP used in the experiment was originally designed for experiments with $\lambda=532$ nm. The actual step height of the SPP is 1596 nm, and over a 25 mm aperture the form accuracy is $\pm$0.5 $\mu$m, and the finish is $\pm$15 nm. % Therefore the SPP induced OAM of $\ell=7$ when used in reflection mode with $\lambda=450$ nm light. 

A quarter wave plate (QWP) was placed in front of the SPP in order to induce a polarization flip. Finally a QWP was placed at the output of the Michelson interferometer in order to prepare the two states of Eq.~\ref{Eqn:psiSO}. The orientation of the QWP determined which output state was being prepared. Hence this QWP was placed on a rotation stage. A lens with f $=150$ mm was used to focus the beam onto the retina of the participants. Several lenses (f= 75, 100, 150, 200, 250, 400, 500 mm) were tested by authors D. S. and C. K. who determined based on their subjective perceptions of the structured light that f=150 mm was the optimal choice.

The mirror was placed on a translation stage in order to induce a controlled phase shift and hence effectively rotate the polarization profile. By translating the mirror along the beam propagation direction we varied $\theta(t)$ by $\approx2\pi/7$ rad$/$s. This is analogous to rotating the polarization direction of a beam to induce high clarity Haidinger's brush~\cite{coren1971use}.

The participants covered their non-viewing eye with an eye patch. The headrest included a chin rest with a variable height and a forehead rest bar. The location of the user lens was optimized for each participant. 

Fig.~\ref{fig:setup}i shows the camera imaged intensity profiles that were observed. The OAM $=-7$ inducing 9 fringes was termed ``many'' while the OAM $=7$ inducing 5 fringes was termed ``fewer''. In the study the participants only observed beams shown in Fig.~\ref{fig:setup}i, and the red circles bound the area ($\approx2\degree$ in field of vision) with good intensity and high quality phase structure that the participants were instructed to observe.

\subsection{Participants}
Experimental participants were recruited from the Institute for Quantum Computing and the School of Optometry and Vision Science at the University of Waterloo. The complete study involved two experimental sessions. Participation required written informed consent and all participants received CAD\$15 per session in appreciation for their time. All research procedures received approval from the University of Waterloo Office of Research Ethics and all participants were treated in accordance with the Declaration of Helsinki.

A total of 12 participants were recruited. Of these, 2 participants did not complete the study. One participant voluntarily withdrew after reporting that they saw many floating features which obscured the stimulus during the familiarization period. The second participant reported discomfort and so they were immediately removed from the study. Therefore, 10 participants completed the experiment.

\subsection{Psychophysical Procedure}
Participants were tested on a psychophysical discrimination task over two experimental sessions. A familiarization period occurred during Session 1 whereby the participants viewed the ``many'' beam while the mirror in the setup was translated, inducing a rotation of the pattern either clockwise or counterclockwise. Participants were asked to observe the region bounded by the red circles in Fig.~\ref{fig:setup}i and indicate the direction of rotation. Participants began the main experiment after five consecutive correct answers in the familiarization task.

After familiarization, participants performed the main psychophysical task. A 5 min dark adaptation period occurred at the start of each session. All participants observed the beam with their preferred eye and the other eye was patched. Each session was composed of 5 blocks with 10 trials each. The trials were separated by $\approx5$ sec, and no break occurred between blocks. At the start of a block, participants observed two alternating presentations of the ``many'' and ``fewer'' beams, each lasting up to 10 seconds. The correct label for each beam was told to the participants. After completing the alternating presentations, participants performed the discrimination task. For each trial a Python 3.6 random number generator was used to determine which state the participant would view. Each trial was presented for no more than 15 seconds (excluding the instances where the participant wished to adjust their position), and the participant verbally indicated the perceived trial type. C. K. was in charge of initializing the QWP orientation via the motorized stage, and he provided the real-time corrective feedback to the participant after each trial. D. S., who did not know the orientation of the QWP in the trials, was present to answer any questions that the participant might have during the study.  In total, each participant completed 100 trials across 2 testing sessions (5 blocks $\times$ 10 trials per block $\times$ 2 sessions).

%\subsection{Data Analysis}
%There was no statistical difference between the results of session 1 and session 2, and therefore the data from both sessions were collapsed for the main analysis. 

%Because a response bias was apparent in the data, the measure of sensitivity $d^\prime$ and the measure of response bias $c$ was calculated from each participant's raw percent correct data. Percent correct is influenced by both a participant's ability to perform the task and the participant's response bias. However, $d^\prime$ is independent of response bias and is therefore a more accurate measure of performance when response bias is present~\cite{wickens2002elementary}. The $d^\prime$ and c scores from all 10 participants were analyzed using two-tailed, one-sample t-tests with 9 degrees of freedom against the null value of 0.

\end{document}